\documentstyle[12pt]{article}

\begin{document}

\begin{titlepage}

\begin{center}

\hfill  CERN-TH/95-124 \\
\hfill  hep-th/9505135\\

\vskip .7in

{\bf FROM QUANTUM MONODROMY TO DUALITY}

\vskip .3in

C\'esar G\'omez$\footnote{D\'epartement de Physique Th\'eorique,
Universit\'e de Gen\`eve, Gen\`eve 6,
Switzerland.}^{\mbox{\footnotesize{,2}}}$ and Esperanza
L\'opez\footnote{Permanent address: Instituto de Matem\'aticas y
F\'isica Fundamental (C.S.I.C.), Serrano 123, 28006 Madrid, Spain.}

\vskip .3in
{\em TH-Division, CERN,}

{\em CH-1211 Geneva 23, Switzerland}

\end{center}

\vskip .3in

\begin{center} {\bf Abstract}  \end{center}
For $N\!=\!2$ SUSY theories with non-vanishing $\beta$-function and
one-dimensional quantum moduli, we study the representation on the
special coordinates of the group of motions on the quantum moduli
defined by $\Gamma_W\!=\!Sl(2;Z)\!/\!\Gamma_M$, with $\Gamma_M$ the
quantum monodromy group.
$\Gamma_W$ contains both the global symmetries and the strong-weak
coupling duality. The action of $\Gamma_W$ on the special coordinates
is not part of the symplectic group $Sl(2;Z)$. After coupling to
gravity, namely in the context of non-rigid special geometry, we can
define the action of $\Gamma_W$ as part of $Sp(4;Z)$. To do this
requires singular gauge transformations on the "scalar" component of
the graviphoton field. In terms of these singular gauge
transformations the topological obstruction to strong-weak duality
can be interpreted as a $\sigma$-model anomaly, indicating the
possible dynamical role of the dilaton field in $S$-duality.
\vskip 2in

\noindent CERN-TH/95-124

\noindent May 1995

\end{titlepage}

{\it 1. Introduction.} Given an $N\!=\!2$ supersymmetric gauge
theory, the geometry of the moduli parametrizing the different vacuum
expectation values allowed by the flat potential, as a consequence of
the non-renormalization theorems \cite{S}, is determined by all the
quantum corrections. The exact quantum moduli for $N\!=\!2$, $SU(2)$
pure Yang-Mills, was first obtained by Seiberg and Witten in
reference \cite{SW1}, extended to $N\!=\!2$ SQCD-$SU(2)$ with $N_f \!
\leq \!4$ in \cite{SW2} and to $N\!=\!2$ $SU(N_c)$ pure Yang-Mills in
references \cite{LY,A}. In all these solutions a beautiful
geometrical picture emerges. Namely associated with the
four-dimensional theory there exists a hyperelliptic curve
$\Sigma_U$, of genus $r$ equal to the rank of the gauge group,
parametrized by the quantum moduli, whose points we denote as
$U\!=\!(u_1,...u_r)$.

For $N\!=\!2$ supersymmetry the geometry of the quantum moduli is
forced to be rigid special K\"ahler \cite{RSK}, which implies, for a
gauge group of rank $r$, the existence of $2r$ holomorphic sections
$(a_i (U) , a_{D i} (U))$ $i\!=\!1,...,r$ of the $Sl(2r;Z)$ bundle
defined by the first homology group $H^1(\Sigma_U;C)$ of the curve
$\Sigma_U$. The physical spectrum is given by the mass formula
\begin{eqnarray}
M &=& \sqrt{2} |Z| \nonumber \\
Z&=& \sum_{i=1}^r (n_{i}^e a_i (U)+ n_{i}^m a_{D i}(U))
\label{1}
\end{eqnarray}
where $n_{i}^e$ and $n_{i}^m$ are the electric and magnetic charges
respectively, and the sections $(a_i (U) , a_{D i} (U))$ can be
represented as
the periods of some meromorphic 1-form $\lambda$ over a basis of
1-cycles $\gamma_i, {\tilde \gamma}_i$
\begin{equation}
a_i = \oint_{\gamma_i} \lambda \;\;\;, \hspace{1cm}
a_{D i} = \oint_{{\tilde \gamma}_i} \lambda
\label{2}
\end{equation}
The mass formula (\ref{1},\ref{2}) already implies that when the
curve degenerates some particle in the spectrum can become massless.

Reducing ourselves to the elliptic case, the metric on the quantum
moduli, given by the rigid special K\"ahler relation
\begin{equation}
\tau(u) = \frac{da_D /du}{da/du}
\label{3}
\end{equation}
turns out to be the elliptic modulus of the curve $\Sigma_u$. The
function $\tau(u)$ defined by (\ref{3}) is the F-term of the low
energy lagrangian and can therefore be used to define the wilsonian
effective coupling and $\theta$-parameter as follows
\begin{equation}
\tau(u) = i\frac{4\pi}{g_{eff}^2(u)} + \frac{\theta_{eff}(u)}{2\pi}
\label{8}
\end{equation}

Being $\tau(u)$ the modulus of an elliptic curve, the positivity of
the coupling constant is automatically assured. Moreover the
Montonen-Olive \cite{MO} duality transformations
\begin{equation}
\tau \rightarrow \frac{a\tau +b}{c\tau +d}  \;\; , \;\;\;\;\;
 \left( \begin{array}{cc}
        a & b \\
        c & d
        \end{array} \right) \in Sl(2;Z)
\label{5}
\end{equation}
coincide with the modular group of the elliptic curve. Defining the
curve $\Sigma_u$ by the vanishing locus of a cubic polynomial in
$P^2$
\begin{equation}
W(x,y,z;u)=0
\label{4}
\end{equation}
the modular group $Sl(2;Z)$, of the elliptic curve defined by
(\ref{4}), appears naturally decomposed into two pieces\footnote{The
analysis we are using here is the algebraic approach to the moduli
problem. This approach is familiar in the study of mirror symmetry,
see for instance \cite{AA}. In that case $\tau(u)$ will have the
meaning of the mirror map.} : {\it i)} the group
$\Gamma_M$\footnote{The monodromy group $\Gamma_M$ is the monodromy
group of the Picard-Fuchs equation for the cycles of the curve
$\Sigma_u$ \cite{FC}.} of monodromy transformations around the
singularities in the $u$-plane, and {\it ii)} the group $\Gamma_W$ of
the coordinate transformations satisfying
\begin{equation}
W(x',y',z';u')=f(u) W(x,y,z;u)
\label{6}
\end{equation}
i.e. transformations on the "target" coordinates which can be, up to
a global factor, compensated by a change in the quantum moduli
coordinate $u$. The explicit relation between $\Gamma_M$, $\Gamma_W$
and $Sl(2;Z)$, which is known in the context of Landau-Ginzburg
theories \cite{L}, is
\begin{equation}
\Gamma_W= \frac{Sl(2;Z)}{\Gamma_M}
\end{equation}

For generic $N\!=\!2$ theories the Montonen-Olive duality (\ref{5})
is lost, mainly because the $\beta$-function is non-vanishing and
that electrically and magnetically charged particles transform in
different representations under supersymmetry. Nevertheless it was
shown in \cite{SW1,SW2} that the monodromy subgroup $\Gamma_M$ of the
$Sl(2;Z)$ transformations (\ref{5}), is actually an exact symmetry of
the quantum theory. This is in general a non-perturbative symmetry if
the monodromy subgroup, as is the case for the examples in
\cite{SW1,SW2}, contains elements with entry $c\! \neq \!0$. The fact
that the $N\!=\!2$ theory is only dual with respect to the monodromy
subgroup means, in particular, that the four-dimensional physics
depends not only on the moduli of the curve but also on its geometry.
This is quite different to what we are used to find in string theory,
where the string only feels the moduli (complex or K\"ahler) of the
target space\footnote{Notice that if we consider, following reference
\cite{LY}, the formal type II string whose target space is defined by
multiplying the algebraic curves defining the quantum moduli, then
for this string, $Sl(2;Z)$ will be its target space duality and it
will contain both $\Gamma_M$ and $\Gamma_W$.}, the difference being
the non-vanishing $\beta$-function for the $N\!=\!2$ theory.

\vspace{1cm}

{\it 2. The meaning of $\Gamma_W$.} On the quantum moduli is also
defined the action of the global $U(1)_{\cal R}$-symmetries which are
broken to some discrete group by instanton effects, so $Z_2$ for pure
$SU(2)$ Yang-Mills and $Z_3$, $Z_2$ for massless SQCD with
$N_f\!=\!1,2$ respectively. These global symmetries are automatically
part of the group $\Gamma_W$.

Each element $\gamma \! \in \! \Gamma_W$ is acting on the quantum
moduli by $\gamma(u)\!=\!u'$, where $u$ and $u'$ are determined by
equation (\ref{6}). It is clear from the definition of $\Gamma_W$
that $\gamma(p_i)\!=\!p_j$ for $p_i$, $p_j$ singular points in the
$u$-plane. The role of $\Gamma_W$ in the characterization of the
monodromies around the different singularities is as follows. For
each singular point $p_i$ we can choose {\it local} special
coordinates in such a way that in the neighbourhood of the
singularity, $a_D(u)$ is determined by the one-loop contribution of
the particles becoming massless at that singular point. The monodromy
of the so defined $a_D(u)$ function will be $T^{k_i}$, for some $k_i$
depending on the quantum numbers of the particles that become
massless at that singular point. Now we can look for the element
$\gamma_i \! \in \! \Gamma_W$ such that $\gamma(\infty )\!=\!p_i$.
Then the monodromy $M_i$ around $p_i$ will be given by
\begin{equation}
M_i= \Gamma_{\gamma_i} T^{k_i} \Gamma_{\gamma_i}^{-1}
\label{16}
\end{equation}
where
\begin{equation}
\tau (\gamma_i(u)) = \frac{a\tau(u) +b}{c\tau(u) +d}  \;\; ,
\;\;\;\;\;
\Gamma_{\gamma_i} =
\left( \begin{array}{cc}
        a & b \\
        c & d
        \end{array} \right)
\label{7}
\end{equation}
Notice that (\ref{7}) reflects the fact that two points $u$, $u'$
related by any element in $\Gamma_W$ correspond to the same complex
structure of the curve $\Sigma_u$. Equation (\ref{16}) clarifies the
physical meaning of $\Gamma_W$, at least the part of $\Gamma_W$ which
maps the singularity at $\infty$ into the rest of the singularities.
In fact this part of $\Gamma_W$ relates the local weak coupling
description around the singular points $p_i$ with the original
coordinates used in the description of the asymptotically free weak
coupling regime at $\infty$. Therefore they play the role  of
defining the dual weak coupling description of a naturally strong
coupling regime. This form of duality is crucial when we want to
argue that the monodromy subgroup is actually an exact symmetry. In
fact in the appropriate dual variables, the monodromy is always a
$T$-transformation, which only changes $\Theta_{eff}/ 2 \pi$ by an
integer number.
We observe, in consequence, that the curve $\Sigma_u$ contains in a
natural and unified way both the information about the
non-perturbative symmetries of the physical system $\Gamma_M$, and
about its dual strong-weak coupling descriptions, enclosed in
$\Gamma_W$.

\vspace{1cm}

{\it 3. The action of $\Gamma_W$ on the special coordinates.} Let us
now consider more closely why the Montonen-Olive duality (\ref{5}) is
actually broken to $\Gamma_M$. This will be another way to see the
dependence of the
four-dimensional physics on the geometry of the curve $\Sigma_u$ and
not only on its moduli. In order to do that, we will consider an
element  $\gamma_i\! \in \! \Gamma_W$ such that
\begin{equation}
\tau(\gamma_i(u)) = -\frac{1}{\tau(u)}
\label{9}
\end{equation}
and we will compare $a(\gamma_i(u))$ with $a_D(u)$. More precisely we
will lift to the bundle the action of $\Gamma_W$ on the $u$-plane.
Notice that  if $a(\gamma_i(u))\!=\!a_D(u)$ we would get a strictly
strong-weak duality, namely that the physics in $\gamma_i(u)$ is dual
to the physics in $u$, where $\gamma_i(u)$ and $u$ correspond, by
(\ref{8}) and (\ref{9}), to strong and weak coupling regimes
respectively.

We will consider first the case of pure $SU(2)$ Yang-Mills. The exact
solution for the quantum moduli is given by the elliptic curve
\cite{SW1}
\begin{equation}
y^2= (x+\Lambda^2)(x-\Lambda^2)(x-u)
\end{equation}
which becomes singular at $u\!=\! \pm \Lambda^2,\infty$, and where
$\Lambda$ is the dynamically generated scale. The monodromy around
the singularities generates the group $\Gamma_2$ of unimodular
matrices congruent to $1$ modulo $2$. The group $\Gamma_W$ in this
case is the dihedral group of six elements \cite{AG} ($[Sl(2;Z)
\!:\!\Gamma_2]\!=\!6$), which on the $u$-plane interchanges the three
singularities. Let us consider $\Lambda^2\!=\!1$ in order to simplify
notation, then $\Gamma_W$ is given by
\begin{equation}
\begin{array}{ccc}
u & \rightarrow & -u \nonumber \\
u & \rightarrow & \frac{u+3}{u-1}
\end{array}
\label{10}
\end{equation}

The transformation (\ref{10}.1) is the part of $\Gamma_W$
corresponding to the global $U(1)_{\cal R}$-symmetry spontaneously
broken to $Z_2$, while the transformation (\ref{10}.2) maps the
singularity at $\infty$ into the point $u\!=\!1$.
The fact that any two points of the quantum moduli related by an
element in $\Gamma_W$ correspond to the same complex structure of the
curve $\Sigma_u$,
together with the definitions (\ref{2}), (\ref{3}) of the sections
$(a, a_D)$ as periods of some 1-form $\lambda$ and of the elliptic
modulus $\tau$, imply that
\begin{equation}
\frac{d\lambda}{du} \sim \lambda_1
\label{13}
\end{equation}
with the proportionality factor determined by the asymptotic
behaviour and where $\lambda_1\!=\!dx\!/\!y$ is the unique everywhere
non-zero holomorphic 1-form, in terms of which the parameter $\tau$
is represented by
\begin{equation}
\tau=\frac{b_2}{b_1} \;\;\;, \hspace{1cm} b_i=\oint_{\gamma_i}
\lambda_1\;\;, \;\;\;i=1,2
\end{equation}
The solution \cite{SW1} for $\lambda$ is
\begin{equation}
\lambda= \frac{\sqrt2}{2 \pi} ( \lambda_2 - u \lambda_1)
\label{97}
\end{equation}
with $\lambda_2\!=\!xdx\!/\!y$. Defining
$u'\!=\!\gamma(u)\!=\!\frac{u+3}{u-1}$ and $x'(x,u)$ as in (\ref{6}),
we
observe that
\begin{equation}
\tau(\gamma(u)) = -\frac{1}{\tau(u)}
\label{11}
\end{equation}
where we have used
\begin{equation}
\lambda(x',\gamma(u)) \equiv \lambda^{\gamma} (x,u) =
\frac{\sqrt2}{2\pi} \left( \frac{2}{1-u}\right)^{1/2} \frac{dx
\sqrt{x+1}}{\sqrt{(x-1)(x-u)}}
\label{35}
\end{equation}
The lift of the action of $\gamma$ on the holomorphic sections is
then given by
\begin{equation}
\begin{array}{llll}
a(\gamma(u)) \;\:= &\oint_{\gamma_{1'}} \lambda^{\gamma} (u)\; = &
- \left( \frac{2}{1-u}\right)^{1/2} \left( a_D(u) + \frac{u+1}{\sqrt2
\pi} b_2(u) \right)  \equiv & a_{D}^{\gamma}(u)  \\
a_D(\gamma(u)) = &\oint_{\gamma_{2'}} \lambda^{\gamma} (u) = &
- \left( \frac{2}{1-u}\right)^{1/2} \left( a(u) + \frac{u+1}{\sqrt2
\pi} b_1(u) \right)\; \; \equiv & a^{\gamma}(u)
\end{array}
\label{12}
\end{equation}
The coordinate transformation $x'(x,u)$ interchanges the cycles
$\gamma_i$ and reverses their orientation, therefore
$\gamma_{1'}\!=\!-\!\gamma_2$ and $\gamma_{2'}\!=\!-\!\gamma_1$.
However, in spite of this and equation (\ref{11}), $a(\gamma(u))$ is
not equal to $a_D(u)$. This is the mathematical manifestation of the
failure of the full Montonen-Olive duality for $N\!=\!2$ theories
with a non-vanishing $\beta$-function. Moreover (\ref{12}) is not
even a symplectic change of special coordinates.

The previous description can be easily generalized to an arbitrary
elliptic curve. In general, for each $\gamma\! \in \! \Gamma_W$ we
get
\begin{equation}
\frac{d\lambda^{\gamma}}{du} = {\tilde f}_{\gamma} (u) \lambda_1
\label{15}
\end{equation}
with
\begin{equation}
{\tilde f}_{\gamma} (u) = l(u) \: \frac{d\gamma(u)}{du} f_{\gamma}(u)
\label{20}
\end{equation}
where $f_{\gamma}(u)$ is determined by
\begin{equation}
\frac{d\lambda_1}{du} = f_{\gamma}(u) \lambda_1
\end{equation}
and $l$ is the proportionality factor in (\ref{13}) associated to the
1-form $\lambda$ which gives the correct physical asympotic
behaviour. For instance, $l\!=\!-\!{\sqrt 2}/4 \pi$ for the 1-form
(\ref{97}).

{}From the above expressions, we obtain
\begin{eqnarray}
\left( \begin{array}{c}
       a_D \\
       a
       \end{array} \right) (\gamma(u)) &=&
\left( \begin{array}{c}
       \oint_{\gamma_{1'}} \lambda^{\gamma} \\
       \oint_{\gamma_{2'}} \lambda^{\gamma}
       \end{array} \right) (u) =  \label{14} \\
&=&
g_{\gamma} (u) \Gamma_{\gamma} \left[
\left( \begin{array}{c}
       a_D \\
       a
       \end{array} \right) (u) + h_{\gamma}(u) \left(
\begin{array}{c}
                                                      b_2 \\
                                                      b_1
                                                      \end{array}
\right) (u)
\right] \equiv \Gamma_{\gamma} \left( \begin{array}{c}
                                      a_{D}^{\gamma} \\
                                      a^{\gamma}
                                      \end{array} \right) (u)
\nonumber
\end{eqnarray}
with $\Gamma_{\gamma}$ defined by (\ref{7}) and where $g_{\gamma}(u)$
and $h_{\gamma}(u)$ are determined by (\ref{15}) in terms of ${\tilde
f}_{\gamma} (u)$. In particular it is easy to see that
$f_{\gamma}\!=\!g_{\gamma}^{-1}$.

The special property of the elements $\gamma\!\in \!\Gamma_W$
corresponding to global symmetries is that for them
$a^{\gamma}\!=\!a$ and $a_{D}^{\gamma}\!=\!a_D$, i.e. they lift to
the bundle as good transformations in $Sl(2;Z)$\footnote{As was
already pointed out in \cite{SW1}, equation (\ref{2}) defines the
sections $(a,a_D)$ up to a sign. This ambiguity appears explicitly
when we express the elliptic modulus $\tau(\gamma(u))$ in terms of
$\tau(u)$. If we strictly use (\ref{2}) and (\ref{35}) we get that,
while $\tau(u)\!\in\!H^+$, $\tau(\gamma(u))\!\in\!H^-$, where
$H^{\pm}$ are respectively the upper and lower half complex plane
\begin{equation}
\tau \rightarrow \frac{a'\tau +b'}{c'\tau +d'}  \;\; , \;\;\;\;\;
 \Gamma_{\gamma}' =\left( \begin{array}{cc}
        a' & b' \\
        c' & d'
        \end{array} \right)
\end{equation}
The matrix $\Gamma_{\gamma}'$ does not belong to $Sl(2;Z)$, but
satisfies $(\Gamma_{\gamma}')^2\!=\!1$ as a consequence of the fact
that $\gamma$ permutes two singularities.
In order to recover a positive $\tau(\gamma(u))$, $a$ or $a_D$ should
be redefined in a sign. Then equation (\ref{7}) is verified,
$\Gamma_{\gamma}\!\in\!Sl(2;Z)$ being the matrix appearing in
(\ref{14}). Notice that this is already evident from (\ref{12}),
where we obtain $a\!\rightarrow\!a_D$, $a_D\!\rightarrow\!a$ which
differs in a sign from an $S$ transformation.}.
Before going into a more detailed analysis of the transformations
(\ref{14}), let us make the following comment on the interplay
between strong-weak coupling duality and scale invariance. We will
consider as an example $SU(2)$ SQCD with $N_f\!=\!1$. In the massless
case there exists three singularities $(p_1,p_2,p_3)$ related by a
global $Z_3$ symmetry \cite{SW2}. When a mass term for the quark is
added, one of these singularities, let us say $p_1$, moves
continuously with the mass to $\infty$, while the others become the
singularities of the $N_f\!=\!0$ theory. For finite mass, the element
$\gamma\!\in\!\Gamma_W$ relating $p_1$ with the singularities
$(p_2,p_3)$ transforms the holomorphic sections $(a,a_D)$ in the way
described by (\ref{14}) (plus a constant shift due to the fact that,
when a mass term is present, the 1-form $\lambda$ has non-vanishing
residues \cite{SW2}).
Geometrically this just means that the monodromy $M_1$ around $p_1$
will not be $T$-conjugated of the monodromies $M_2$ or $M_3$. What
this teaches us is that a finite mass breaks the global $U(1)_{\cal
R}$ symmetries, as they are represented in $\Gamma_W$, in formally
the same way as the non-vanishing $\beta$-function, i.e. a non-zero
scale $\Lambda$, breaks the Montonen-Olive duality, namely inducing
on the holomorphic sections $(a, a_D)$ changes of the type (\ref{14})
with $h_{\gamma}\!\neq\!0$. This fact strongly indicates that, at
least for SUSY gauge theories, a necessary condition for duality will
be to have, in addition to scale invariance, a non-anomalous
$U(1)_{\cal R}$-symmetry\footnote{The existence of non-anomalous
$U(1)_{\cal R}$ symmetries together with scale invariance implies
that even for $N\!=\!1$ SUSY theories, the conformal phase shares
many aspects of $N\!=\!2$ theories. This fact is crucial in the
$N\!=\!1$ duality between $SU(N_c)$ and $SU(N_f\!-\!N_c)$ with $N_f$
quarks \cite{SS}.}. In a different language, equation (\ref{14})
reflects the dependence of the four dimensional physics on the
geometry of the elliptic curve, i.e. the way it changes for two
points $u,\gamma(u)$ which describe the same moduli.

\vspace{1cm}

{\it 4. Coupling to gravity.} A natural way to try to make sense of
equation (\ref{14}) is considering the coupling to gravity.
Intuitively we can think of (\ref{14}) as a $Sp(4;Z)$ transformation
by interpreting the extra piece in the periods $b_i$, as
contributions from the gravitational sector associated to the
additional $U(1)$ field present in $N\!=\!2$ supergravity: the
graviphoton.
Due to the presence of the graviphoton, it is necessary to introduce
a new
(non-dynamical) special coordinate $(a_0, a_{D 0})$
and to define the special manifold (quantum moduli) projectively. In
this picture the transformation from $u$ to $\gamma(u)$,
$\gamma\!\in\!\Gamma_W$, will become a good element in $Sp(4;Z)$ if
at the same
time we perform a, in general singular, $U(1)$ gauge transformation
of the K\"ahler-Hodge line bundle which we have naturally defined
when we pass from rigid to non-rigid special
geometry\footnote{Another reason supporting this idea comes from
Landau-Ginzburg theories. For Landau-Ginzburg models, it is possible
to build all gravitational descendant fields inside the matter sector
\cite{Lo}. This allows us to interpret the reparametrizations of the
superpotential $W$ as contributions from gravitational descendants
\cite{K,U}. Therefore transformations (\ref{14}) should admit a
natural representation when gravity is turned on, i.e. they should be
elements of $Sp(4;Z)$.}\cite{SG}.

More precisely, denoting $V\!=\!(a,a_D)$, the rigid special geometry
for a 1-dimensional moduli space, is defined by
\begin{equation}
\begin{array}{lll}
d_u V &= & U \\
D_u U &=& C_{uuu} G_{u {\bar u}}^{-1} {\bar U} \\
d_u {\bar U} &=& 0
\end{array}
\label{17}
\end{equation}
where $G_{u {\bar u}}\!=\! {\mbox Im} \tau(u)$ is the metric over the
moduli space, the Yukawa coupling $C_{uuu}$ is given by
\begin{equation}
C_{uuu} = \frac{d\tau}{du}
\label{18}
\left( \frac{da}{du} \right)^2
\end{equation}
and the covariant derivative is
\begin{equation}
D_u = d_u - \Gamma_u \;\;\;,\hspace{1cm} \Gamma_u=G_{u {\bar u}}^{-1}
(d_u G_{u {\bar u}})
\label{19}
\end{equation}

For each $\gamma\!\in\!\Gamma_W$ we define $V^{\gamma}
\!=\!(a^{\gamma}, a_{D}^{\gamma})$ with $a^{\gamma}$,
$a_{D}^{\gamma}$ given by (\ref{14}).
It is easy to verify that $V^{\gamma}$ satisfy (\ref{17}) with
\begin{equation}
\begin{array}{lll}
D_{u}^{\gamma} &=& D_u - d_u ln \: {\tilde f}_{\gamma} \\
U^{\gamma} &=& {\tilde f}_{\gamma} U
\end{array}
\label{50}
\end{equation}
where ${\tilde f}_{\gamma}$ is given by (\ref{20}) and the quantities
$C_{uuu}^{\gamma},G_{u{\bar u}}^{\gamma}$ are defined according to
(\ref{18}) and  (\ref{19})
in terms of $V^{\gamma}$ instead of $V$.
This modification of $D_u$ is not allowed in rigid special geometry,
which is showing again that changing from $V$ to $V^{\gamma}$ is not
a symplectic transformation. However (\ref{50}) can be naturally
interpreted from the point of view of non rigid special geometry,
whose defining relations are
\begin{equation}
\begin{array}{lll}
D_u V &= & U \\
D_u U &=& e^K C_{uuu} G_{u {\bar u}}^{-1} {\bar U} \\
d_u {\bar U} &=& G_{u{\bar u}} {\bar V} \\
d_u {\bar V} &=& 0
\end{array}
\label{22}
\end{equation}
with $V\!=\!(a_0,a_1,a_{D 1}, a_{D 0})$ and the covariant derivative
\begin{equation}
D_u= d_u - G_{u {\bar u}}^{-1} (d_u G_{u {\bar u}}) + d_u K
\label{23}
\end{equation}
where the piece $d_u K$, $K(u,{\bar u})$ being the K\"ahler
potential, is the $U(1)$ connection associated to the Hodge line
bundle over the quantum moduli, present when we couple to gravity.
Notice that the vector $U$ now acquires a
non-holomorphic part
\begin{equation}
U=d_u V + d_uK \: V
\end{equation}
which is at the origin of the third equation in (\ref{22}).

{}From (\ref{50}) we observe that working with the sections
$(a^{\gamma}, a_{D}^{\gamma})$ amounts, at the level of the vector
$U$, $\bar U$, to multiplying by a global factor. In the framework of
special geometry, this can be interpreted as a change in the
projective coordinate
\begin{equation}
a_0 = 1 \rightarrow a_{0}^{\gamma} = {\tilde f}_{\gamma}
\label{26}
\end{equation}
or, equivalently, as the gauge transformation
\begin{equation}
K \rightarrow K - {\mbox ln} {\tilde f}_{\gamma} - {\mbox ln} {\bar
{\tilde f}_{\gamma}}
\label{27}
\end{equation}
and therefore as a change in the covariant derivative (\ref{23}) of
the form required by (\ref{50}).
In special geometry, we can now define $\lambda^{\gamma}$ by the
equation
\begin{equation}
D_{u}^{\gamma} \lambda^{\gamma}= {\tilde f}_{\gamma} \lambda_1
\label{40}
\end{equation}
with $D_{u}^{\gamma}$ defined by (\ref{50}.1) and (\ref{23}).
Equation (\ref{40}) is motivated by the first relation (\ref{22}). We
now define the sections $a^{\gamma}$ and $a_{D}^{\gamma}$ by the
corresponding integrals around 1-cycles of the solution to
(\ref{40}). They satisfy $a^{\gamma}\!=\!{\tilde f}_{\gamma} a$,
$a_{D}^{\gamma}\!=\!{\tilde f}_{\gamma} a_D$, and therefore the
action of $\gamma$ on the special geometry coordinates is given by
\begin{equation}
\left( \begin{array}{c}
       a_{D 0}^{\gamma}  \\
       a_{D 1}^{\gamma}  \\
       a_{1}^{\gamma}    \\
       a_{0}^{\gamma}
       \end{array} \right) = {\tilde f}_{\gamma}
\left( \begin{array}{cccc}
       1 & 0 & 0 & 0  \\
       0 & a & b & 0  \\
       0 & c & d & 0  \\
       0 & 0 & 0 & 1
       \end{array} \right)
\left( \begin{array}{c}
       a_{D 0}\\
       a_{D 1}\\
       a_1    \\
       a_0
       \end{array} \right)  \nonumber  \;\;\;, \;\;\;\;
\left( \begin{array}{cc}
       a & b \\
       c & d
       \end{array} \right) = \Gamma_{\gamma}
\label{41}
\end{equation}
with $\Gamma_{\gamma}$ given by (\ref{7}). Equation (\ref{41})
implies that the action of $\Gamma_W$ can be represented, once we
couple to gravity, by an element of $Sp(4;Z)$ plus the K\"ahler gauge
transformation (\ref{27}).

Notice that in rigid special geometry, the action of $\gamma$ on
$(a,a_D)$ was defined by the condition
$a^{\gamma}(u)\!=\!a(\gamma(u))$,
$a_{D}^{\gamma}(u)\!=\!a_{D}(\gamma(u))$, which was at the origin of
the
non-symplectic transformations (\ref{14}). It is important to analyse
to what extent this condition is verified by the non-rigid $\gamma$
transformations (\ref{41}).
Let us consider an element $\gamma\!\in\!\Gamma_W$ that interchanges
the singularity at $\infty$ with a finite singular point, say $p_1$,
while leaving the rest fixed. In the case of zero masses for the
quarks, the asymptotic behaviour of the sections $(a,a_D)$ at
$\infty$ is given by \cite{SW1,SW2}
\begin{equation}
a(u')= \frac{1}{2} \sqrt{2u'}\;\;\;,\hspace{1cm}
a_D(u')=i\frac{k_{\infty}}{4\pi} \sqrt{2u'}\: {\mbox ln}
\frac{u'}{\Lambda^2}
\label{71}
\end{equation}
At the singular point $p_1$ some particle in the spectrum becomes
massless. Using the dual description and up to an $Sl(2;Z)$ rotation,
the special coordinates behave as
\begin{equation}
a(u)=c_0 (u-p_1)\;\;\;,\hspace{1cm} a_D(u)=c_1
-\frac{ik_1}{2\pi}a\:{\mbox ln} (u-p_1)
\label{70}
\end{equation}
with $c_0,c_1$ constants.
Comparing the two limits, equation (\ref{14}) implies\footnote{In
fact, from (\ref{14}) we get $g_{\gamma}h_{\gamma}\!\sim\!\sqrt{u'}$
instead of (\ref{60}). However it can be seen that the function
$h_{\gamma}$ is regular at $u=p_1$ and therefore it is enough to
consider (\ref{60}) in the following computations.}
\begin{equation}
g_{\gamma}(u) \sim \sqrt{u'}
\label{60}
\end{equation}
The map $u'\!=\!\gamma(u)$ is, of course, singular at $u\!=\!p_1$.
Taking into account only its singular part, and for a certain
constant $C$, we have
\begin{equation}
u'=C(u-p_1)^{-k}\;\;\;,\hspace{1cm} k>0
\label{61}
\end{equation}
where $k$ is determined again from (\ref{14}), by correctly
reproducing the monodromy at $\infty$
\begin{equation}
k=k_1/k_{\infty}
\label{90}
\end{equation}
Substituting now (\ref{60}), (\ref{61}) in the expression of the
function ${\tilde f}_{\gamma}$, we obtain
\begin{equation}
{\tilde f}_{\gamma}=\frac{du'}{du} g_{\gamma}^{-1}\sim
\frac{\sqrt{u'}}{u-p_1}
\label{91}
\end{equation}
Therefore the special coordinates $a_{1}^{\gamma}, a_{D 1}^{\gamma}$
defined by (\ref{41}) have the expected asymptotic behaviour at
$u\!\rightarrow\!p_1$, namely they tend to $a(\gamma(u)),
a_{D}(\gamma(u))$.
Notice also that, if $V\!=\!(a_0,a_1,a_{D 1},a_{D 0})$ satisfy the
special geometry relations (\ref{22}), so does the vector
$V^{\gamma}$ defined by (\ref{41}),  with $U^{\gamma}\!=\!{\tilde
f}_{\gamma} U$ as we should expect from (\ref{50}.1).

In our previous construction, the extra special coordinate $(a_0,a_{D
0})$ associated with the graviphoton plays a role similiar to that of
the mass in SQCD.
In fact when a mass term is added, there appear monodromies which are
not in $Sp(2;Z)$ $(v \rightarrow Mv + c, M\!\in\!Sl(2;Z))$. These
monodromies have perfect sense once we formally treat the mass as a
field \cite{SW2}. In the case of non-vanishing $\beta$-function we
are trying to give sense to the strong-weak duality transformations
$\gamma\!\in\Gamma_W$ as element in $Sp(4;Z)$ by including as an
extra degree of freedom the graviphoton multiplet.

\vspace{1cm}

{\it 5. Duality and $\sigma$-model anomalies.} From physical grounds
we should expect that if the $N\!=\!2$ theories we are working with
are some low-energy limit of a string theory, then the stringy
effects will be able to restore the whole duality invariance. The
picture that emerges from our previous construction seems to indicate
a possible way to achieve this goal.

In $N\!=\!2$ SUGRA, and this is specially clear when we formulate the
theory starting with conformal supergravity and passing later to
Poincar\'e supergravity, the projective coordinate $a_0$ is not a
real degree of freedom. Equivalently, the chiral $U(1)$ gauge field
$A_{\mu}$ of the Weyl supermultiplet\footnote{The Weyl supermultiplet
appears in the context of conformal gravity. It contains \cite{VP}
the gauge fields associated with the superconformal symmetries,
namely general coordinates and local Lorentz transformations,
dilatations, special conformal boosts and local supersymmetries.
In addition, for $N\!=\!2$, there exists a local chiral $SU(2)$ and
$U(1)$. Notice that this chiral $U(1)$ is the gauge symmetry defining
the K\"ahler-Hodge line bundle of special geometry.} is an auxiliary
field that can be eliminated by solving the constraints in the same
way as we are used to do in non-linear $\sigma$-models. Up to
fermionic terms, the field $A_{\mu}$ can be expressed in terms of the
K\"ahler potential as follows
\begin{equation}
A_{\mu}= \frac{i}{2}(\partial_{\mu} z \left(dz/du\right)^{-1} \!d_u K
+ \partial_{\mu} {\bar z} \left(d{\bar z}/d{\bar u}\right)^{-1}\!
d_{\bar u} K )
\label{49}
\end{equation}
where $z\!=\!a_1\!/\!a_0$ is the homogeneous special coordinate.
The transformation (\ref{27}) over the K\"ahler potential can be
interpreted as  a gauge transformation on $A_{\mu}$, which
corresponds to passing from a coordinate patch in the quantum moduli
space around $u\!=\!p_1$, to a local coordinate patch
$u'\!=\!\gamma(u)$ around $\infty$, for $\gamma\!\in\!\Gamma_W$. This
transformation is characterized by the parameter $k$ in (\ref{61}),
which is different from zero as a consequence that $\Gamma_{\gamma}$
is not in the abelian subgroup generated by $1$ and $T$.

As we have shown in the previous paragraph, once we use non rigid
special coordinates we get for the action of $\Gamma_{\gamma}$ the
representation (\ref{41}), which in particular means that if
$\tau(\gamma(u))\!=\!\frac{-1}{\tau(u)}$ then
$a(\gamma(u))\!=\!{\tilde f}_{\gamma}a_D(u)$, i.e. strong-weak
coupling duality if the special coordinates $a$ and ${\tilde
f}_{\gamma}a$ can be considered as gauge equivalent. The fact that
the gauge transformation (\ref{27}) defined by ${\tilde f}_{\gamma}$
is in general singular can be at the origin of a topological
obstruction to mod by $\Gamma_W$ of the type found in $\sigma$-model
anomalies \cite{SA}. Using the field $A_{\mu}$, equation (\ref{27})
and assuming that the quantum corrections corresponding to
integrating over the fermions have already been taken into account in
the geometry, singularities, of the quantum moduli, we can use, for a
compactified quantum moduli, the following quantity as an indication
of this topological obstruction\footnote{For $\sigma$-model
anomalies, the topological obstruction is defined by $\int_{S^2
\times S^d} {\hat \phi}^{\ast} Ch(T {\cal M})$, with $d$ the
dimension of the space-time and $T {\cal M}$ the tangent bundle to
the $\sigma$-model manifold. The map $\hat \phi$ should define a
non-contractible two-parameter family of $\sigma$-model
configurations. In our case, heuristically, it is the compactified
$u$-plane and their singularities that define the "non-contractible"
two-sphere, determining the reparametrizations and in this way the
configuration of the auxiliary gauge field $A_{\mu}$ on the quantum
moduli.}
\begin{equation}
\nu = \frac{1}{2\pi i} \oint_{C_{\infty}} \!\! d \:{\mbox ln} {\tilde
f}_{\gamma} =
1+ \sum_i \frac{k_i}{2 k_{\infty}}
\label{95}
\end{equation}
where $C_{\infty}$ encircles the singularity at $\infty$ and the
coefficients  $k_{\infty}$, $k_i$ are given in (\ref{71}) and
(\ref{70}) respectively. Equation (\ref{95}) was obtained up to the
normalization factors which can be derived from (\ref{49}).
The sum in (\ref{95}) is over the singular points at which the
function
${\tilde f}_{\gamma}$ has a pole, namely
$\gamma(p_i)\!=\!\infty$\footnote{It is worth recalling that
equations (\ref{60}-\ref{91}) and (\ref{95}) could be different for
an element $\gamma\!\in\!\Gamma_W$ mapping finite singular points
between themselves.}.

The quantity (\ref{95}) is showing the existence of an anomaly to
define the theory on the moduli space of complex structures of the
curve, i.e. to mod by the action of $\Gamma_W$.
Once we have interpreted (\ref{95}) as an anomaly, it will be natural
to look for some compensating WZ term. This is in general not
possible for $\sigma$-model anomalies \cite{SA}. In our case, the
introduction of the dilaton will, very likely, play that role. In
fact, the anomaly (\ref{95}) is heuristically indicating that $a_0$
can't be globally gauged away, which is strongly asking for an extra
scalar degree of freedom in the gravity sector. We hope to address
these problems in more detail elsewhere.

\vspace{1cm}

This work was partially supported by
european community grant ERBCHRXCT920069, by PB 92-1092 and by OFES
contract number 93.0083. The work of E.L. is supported by a
M.E.C. fellowship AP9134090983.

\end{document}